\documentclass[aps,prl,preprint,superscriptaddress,showpacs,amssymb]{revtex4}
\usepackage{graphicx}
\usepackage{graphics}
\usepackage{amsmath}
\usepackage{subfigure}
\usepackage{color}
\usepackage{doi}

\begin{document}

\title{Logarithm corrections in the critical behavior of the Ising model on a triangular lattice modulated with the Fibonacci sequence}
\author{T.F.A. Alves}
\affiliation{Departamento de F\'{\i}sica, Universidade Federal do Piau\'{i}, 57072-970, Teresina - PI, Brazil}
\author{G.A. Alves}
\affiliation{Departamento de F\'{i}sica, Universidade Estadual do Piau\'{i}, 64002-150, Teresina - PI, Brazil}
\author{M.S. Vasconcelos}
\affiliation{Escola de Ci\^encias e Tecnologia, Universidade Federal do Rio Grande do Norte, 59078-900, Natal - RN, Brazil}

\date{Received: date / Revised version: date}

\begin{abstract}

We investigated the critical behavior of the Ising
model in a triangular lattice with ferro and anti-ferromagnetic
interactions modulated by the Fibonacci sequence, by using
finite-size numerical simulations. Specifically, we used a replica
exchange Monte Carlo method, known as Parallel Tempering, to
calculate the thermodynamic quantities of the system. We have
obtained the staggered magnetization $q$, the associated
magnetic susceptibility ($\chi$) and the specific heat $c$, to
characterize the universality class of the system. At the
low-temperature limit, we have obtained a continuous phase
transition with a critical temperature around $T_{c} \approx
1.4116$ for a particular modulation of the lattice according to the
Fibonacci letter sequence. In addition, we have used finite-size scaling relations
with logarithmic corrections to estimate the critical exponents
$\beta$, $\gamma$ and $\nu$, and the correction exponents
$\hat{\beta}$, $\hat{\gamma}$, $\hat{\alpha}$ and $\hat{\lambda}$.
Our results show that the system obeys the Ising model universality
class and that the critical behavior has logarithmic corrections.

\end{abstract}


\maketitle

\section{Introduction}

At the beginning of the last century, with the experiments of
Rutherford \cite{SirErnestRutherford.1914}(awarded with the Nobel
Prize in Chemistry in 1908) on the discovery and interpretation of
the dispersion a beam of alpha particles directed to a fine gold
leaf, there was a great effort in the scientific community in
knowing the structure of matter, especially in its solid state.
Since then, one aspect of the structure of matter that has become
well known has been the {\it symmetry of translation} of atoms
organized into crystals. However, in 1984, in their work entitled
``Metallic Phase with Long-Range Orientational Order and No
Translational Symmetry'', Shechtman et al.
\cite{PhysRevLett.53.1951}, that was awarded the Nobel Prize in
Chemistry in 2011, has shown new solid materials that exhibit a new
symmetry in the structure of matter: the quasicrystals.
Quasicrystals are a particular type of solid structure that can have
unusual discrete point group symmetries, not expected from a
translational symmetric Bravais lattice in two dimensions. It has
also been shown that quasicrystals can have icosahedral symmetry in
three dimensions. It had long been known that icosahedral symmetry
is not allowed for periodic objects like crystals. It is forbidden
in crystallography. However, systems like quasicrystals are not
periodic but exhibit an exotic (forbidden) symmetry. They have a
long-range order called quasiperiodicity, that characterize their
unique and fascinating properties: they follow mathematical rules
\cite{Steinhard.1987}. Some metallic alloys
\cite{Steurer2012,Macia2006}, solf-matter systems
\cite{Fischer2011}, supramolecular dendritic systems
\cite{Zeng2004,Zeng2005}, and copolymers
\cite{Takano2005,Hayashida2007} are examples of quasiperiodic
systems. In common, these systems possess magnetic properties
sensitive to the local atomic structure such as the atomic distance,
coordination number and the kind of the nearest-neighbor atoms. Over
of the last decades, the quasicrystals have to contribute to advance
knowledge about the atomic scale structure. However, some questions
remain open, such as magnetic properties of the physical systems
that present these quasiperiodic structures. For example, a question
unanswered is whether long-range antiferromagnetic (AFM) order can
be sustained in real quasicrystalline systems.

In recent decades, research on quasicrystals has contributed to the
advancement of knowledge about the structure of the matter at the
atomic scale. However, some questions remain open. One is: what are
the magnetic properties of the physical systems that have these
atoms organized in quasi-periodic order? Specifically, another
question unanswered is: a long-range antiferromagnetic (AFM) order
can be sustained in real quasicrystalline systems?

The lack of translational symmetry in the quasiperiodic structure
can induce, as consequence, anomalous properties different from a
regular crystal. In the other hand, the quasicrystals are different
from disordered materials, because they possess self-similar
properties, i.e., any finite section of the structure represent
exactly or approximately the structure of the quasicrystal in a
distance of a degree of its linear (2D or 3D) scale. Also, the
quasicrystal can present long-range correlations in sufficiently low
temperature. If all these ingredients come into place in a simple
lattice model, the physical observables obtained can be affected by
the quasiperiodicity of the lattice itself. Instead of the
antiferromagnetic periodic crystals, the antiferromagnetic
arrangement in quasicrystals has shown a different behavior from
usual crystals. For example, rare earth-containing quasicrystals
\cite{PhysRevB.57.R11047,SCHEFFER2000629} exhibit an aperiodic
ferrimagnet freezing phase at low-temperature.

The theoretical works
\cite{PhysRevLett.92.047202,PhysRevLett.93.076407,Matsuo2004421,PhysRevB.71.115101,PhilosMag.86.733}
suggest the possibility of a non-trivial magnetic ordering for the
quasicrystals. Although no antiferromagnetic quasicrystal has not
yet been discovered, these studies reveal that the behavior of some
quasicrystals at low-temperature show magnetic topological order and
frustration. The debate in question allows us to study new
theoretical models to answer the questions that remain open. Amongst
them, the change of the critical exponents is partially answered by
the Harris-Luck criterion, valid for ferromagnetic
systems\cite{EurophysLett.24.359}.

On another hand, there are ways to control the disorder in a certain
systems and obtain a transition from a long-range order to a
quenched disorder \cite{PhysRevB.61.15738,RevModPhys.71.1125} or,
alternatively, the quasiperiodic order by modifying the exchange
strengths and signals. The types of the quasiperiodic order that can
be used to model the quasicrystals are: 1) we can modulate the
interactions; 2) we can change the geometry of the crystal lattice;
We choose the second option by considering a triangular lattice with
ferro and antiferromagnetic exchange interactions modulated by a
quasiperiodic sequence.

Recently, we published two papers about quasiperiodic models based
on Fibonacci and Octonacci sequence, respectively. The models were
applied in a square lattice. We have shown that is possible obtained
the critical behavior of the two models, being that both have
presented second order transition, with critical temperature,
$T_{c}=1.274$ and $T_{c}=1.413$. The square lattice
was modulated in a way that generates frustrated plaquettes throughout the
lattice. Therefore, depending on the rate of the frustrated
plaquettes, the system is induced to an aperiodic diluted
ferrimagnet phase \cite{PhysRevE.89.042139}. In both the
quasiperiodic models, we found very interesting results. We can
highlight the curves of the heat specific $c$ that are not collapsed
with the inverse of the $\mbox{ln} L$, where $L$ is the size of the
lattice. We had to use the logarithmic scale corrections in order to
achieve the collapse of the specific heat curves \cite{kenna:2012}.
This new theory opened a promising field of research about systems
that can display unusual (quasiperiodic or aperiodic) orderings at
low-temperature.

In this work, we considered the triangular lattice and obtain the
relevant thermodynamic properties of the Ising model in two
dimensions with positive and negative exchange interactions with the
same strength modulated by Fibonacci quasiperiodic sequence. On the
triangular lattice, we have several ways to modulate the lattice and
have some control over the frustration rate and the ratio of the
antiferromagnetic interactions. By controlling the modulation of the
lattice we can investigate the influence of quasiperiodic modulation
in critical properties. In section 2 we present our model. In
section 3 we show the results and discussion. Finally, in section
4we present our conclusions.

\section{Model and Simulations}

We consider the Ising model in a triangular lattice with only first
neighbor interactions. The Hamiltonian of the model is given
by\cite{ZPhys.31.253}
\begin{equation}
\mathcal{H}=-\sum_{\langle i,j \rangle}J_{ij}S_{i}S_{j},
\end{equation}
where $S_{i}$ and $S_{j}$ are the spin on sites $i$ and $j$,
respectively and its values can be $\pm 1$. The exchange
interactions $J_{ij}$ between first neighbor spins $S_{i}$ and
$S_{j}$ are modulated according to an aperiodic letter sequence and
they can have values $1$ and $-1$.

We have used the Fibonacci letters sequence, to investigate if a
particular marginal quasiperiodic order can confirm a new the
universality class or not. The Fibonacci sequence can be obtained
from the substitution rules $A\rightarrow AB$ and $B\rightarrow A$
in 1D\cite{VASCONCELOS.2006,VASCONCELOS.2007,PhysRevE.93.042111}, or
alternatively from the substitution ruler: $S_n=S_{n-1}S_{n-2}$ (for
$n \geq 2$), with, $S_1=A$ and $S_2=AB$. Any generation of the
aperiodic sequence can be constructed from the previous generation
by replacing all letters \textit{A} with \textit{AB} and all letters
\textit{B} with \textit{A}. Starting with the letter \textit{A}, by
repetitive applications of the substitution rule we can obtain the
successive iterations of the Fibonacci sequence in 1D.

\begin{figure}[h!]
\begin{center}
\includegraphics[scale=0.6]{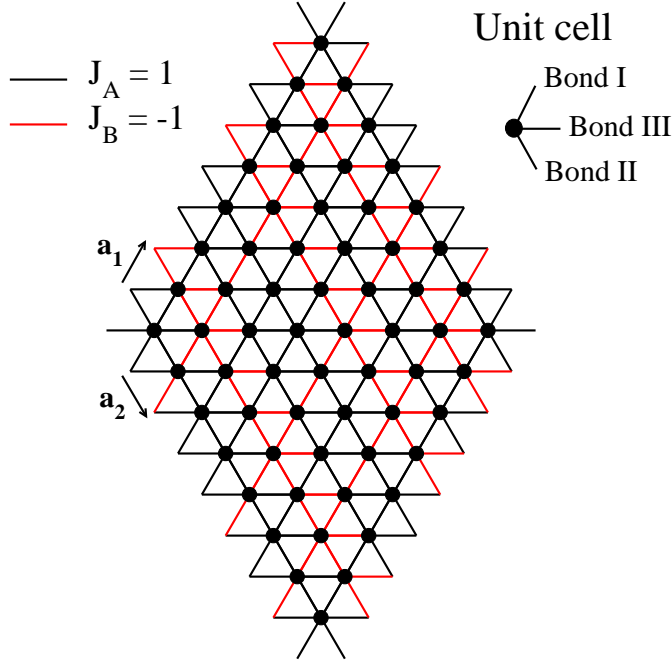}
\end{center}
\caption{(Color online) Example of a lattice with exchange
interactions modulated by the Fibonacci sequence. The black and red
lines stand for exchange interaction strengths $J_{A}=1$
(ferromagnetic) and $J_{B}=-1$ (anti-ferromagnetic) respectively. We
used the Fibonacci letter sequence, which is obtained from the
substitution rules $A\rightarrow AB$ and $B\rightarrow A$ which
means that any generation of the lattice can be constructed from the
previous generation by replacing all letters \textit{A} with
\textit{AB} and all letters \textit{B} with \textit{A}. The
modulation of the bonds between sites of the triangular lattice was
made by means of the Fibonacci sequence at the lattice direction
specified by Bravais vectors $\textbf{a}_{1}$ and $\textbf{a}_{2}$,
according to way C) described in the text.}
\label{fig_quasiperiodic_lattice_1}
\end{figure}

\begin{figure}[t]
\begin{center}
\includegraphics[scale=0.4]{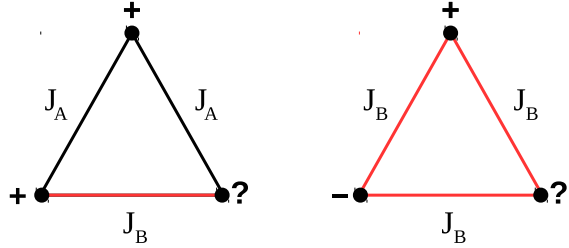}
\end{center}
\caption{Frustrated plaquettes of the triangular lattice. The black
and red lines stand for exchange interaction strengths $J_{A}=1$
(Ferromagnetic) and $J_{B}=-1$ (Anti-ferromagnetic) respectively. If
a given site, represented by the black circle, has an spin up and
interacts with a neighbor site through of a $J_{A}$ exchange
interaction, the spin of this site remains up. Otherwise, if the
interaction is through of a $J_{B}$ exchange, the spin orientation
of the neighboring site changes to down for minimal energy. For the
two possibilities of a frustrated plaquette in the triangular
lattice, the spin orientation on right corner can be either up or
down for minimal energy.} \label{fig_quasiperiodic_lattice}
\end{figure}

The modulation of the exchange interactions in the triangular
lattice was made by means of the Fibonacci sequence. We consider the
three bonds present in a unitary cell, named I, II, and III, which
connects one site $i$ at the position $\mathbf{r}_{i}$ and the
neighbors placed on positions I) $\mathbf{r}_{i}+\mathbf{a}_{1}$,
II) $\mathbf{r}_{i}+\mathbf{a}_{2}$ and III)
$\mathbf{r}_{i}+\mathbf{a}_{1}+\mathbf{a}_{2}$, where
$\mathbf{a}_{j}$ are the Bravais vectors of the lattice, given by
\begin{eqnarray}
\mathbf{a}_{1} &=& \frac{1}{2}\mathbf{i} + \frac{\sqrt{3}}{2}\mathbf{j} \nonumber \\
\mathbf{a}_{2} &=& \frac{1}{2}\mathbf{i} - \frac{\sqrt{3}}{2}\mathbf{j}.
\end{eqnarray}
The modulation can be done in three ways in order to generate plaquette frustration,
according to the unitary cell bonds and lattice directions. Lattice bonds
are sorted as
\begin{enumerate}
\item[I)]   Bonds I)   are lattice bonds at $\mathbf{a}_{1}$ direction;
\item[II)]  Bonds II)  are lattice bonds at $\mathbf{a}_{2}$ direction;
\item[III)] Bonds III) are lattice bonds at $\mathbf{a}_{1}+\mathbf{a}_{2}$ direction.
\end{enumerate}
A modulation can be done by changing bonds I), II) and III) at one
defined lattice direction according to the Fibonacci letter sequence.
The three modulation schemes considered here are
\begin{enumerate}
\item[A)]       Modulating bond  I)               at the lattice line  defined
by $\mathbf{a}_{2}$ direction;
\item[B)]       Modulating bonds I) and II)       at the lattice lines defined
by $\mathbf{a}_{2}$ and $\mathbf{a}_{1}$ directions, respectively;
\item[C)]       Modulating bonds I), II) and III) at the lattice lines defined
by the $\mathbf{a}_{2}$, $\mathbf{a}_{1}$ and $\mathbf{a}_{1}$ directions, respectively.
\end{enumerate}
We performed simulations in the triangular lattice for each one of
the modulation schemes A), B) and C) with the Fibonacci sequence in
order to characterize the magnetic ordering of the spins at low
temperature. We show an example of such lattice modulated according
to way C) in Fig.(\ref{fig_quasiperiodic_lattice_1}). The Fig.
(\ref{fig_quasiperiodic_lattice}) displays the possibilities we will
have a plaquette frustrated throughout the lattice. In the
triangular lattice, each plaquette is composed of the three sites,
the links are made by $J_{ij}$ exchange interaction strength. The
plaquette frustrates if there are two bonds with $J_{A}=1$ and one
bond $J_{B}=-1$ or the three bonds with $J_{B}=-1$. We focused on
modulation C) because of its distinctive critical behavior as we
present in next section.

By using the Replica Exchange Monte Carlo technique (also known as
Parallel Tempering)\cite{PhysChemChemPhys.7.3910,
PhysRevLett.57.2607,JPSJ.65.1604,Geyer.1991}, which is suited to
find the ground state of such systems with alternating interactions,
we obtained the staggered magnetization order parameter $\langle q
\rangle$, the associated susceptibility $\chi$, the specific heat
$c$ and Binder cumulant $g$
\begin{eqnarray}
q & = & \frac{1}{N}\sum_{i}^{N} S^{0}_{i}S_{i} \label{s0s_parameter}\\
\chi & = & N\left( \langle q^{2} \rangle - \langle q \rangle^{2} \right)/T, \label{susceptibility}\\
c & = & N\left(\langle \mathcal{H}^{2} \rangle - \langle \mathcal{H} \rangle^{2}\right)/T^{2}, \label{specific_heat} \\
g & = & 1-\frac{\langle q^{4} \rangle}{3\langle q^{2} \rangle^{2}}, \label{binder_cumulant}
\end{eqnarray}
where $\langle...\rangle$ stands for a thermal average over
sufficiently many independent steady state system configurations,
q is the staggered magnetization of the system, corresponding to a
ferrimagnet phase where $S^{0}_{i,j}$ is the ground state,
and $L$ and $T$ are the lattice size and the absolute temperature,
respectively. We used the following values of the lattice size
$L$: 34, 55, 89, 144 and 233, which are Fibonacci's numbers $F_n$,
given by the recursion rule:
\begin{equation}
F_n = F_{n-2} + F_{n-1}, \label{pellrecursionrule}
\end{equation}
where $P_{0}=1$ and $P_{1}=1$. The total number of spins for each lattice size is $N=L^{2}$.

To determine the critical behavior, we have used the following
Finite Size Scaling (FSS) relations\cite{PhysA.391.1753}, with
logarithmic
corrections\cite{PhysRevLett.96.115701,PhysRevLett.97.155702,PhysRevE.82.011145,kenna:2012}
\begin{eqnarray}
q & \propto & L^{-\beta/\nu}\left(\ln{ L }\right)^{-\hat{\beta}-\beta\hat{\lambda}}f_q(\vartheta), \label{q_fss} \\
\chi & \propto & L^{\gamma/\nu}\left(\ln{ L }\right)^{-\hat{\gamma}+\gamma\hat{\lambda}}f_{\chi}(\vartheta), \label{susceptibility_fss} \\
c & \propto & \left( \ln L \right)^{\hat{\alpha}}f_{c}(\vartheta), \label{specificheat_fss} \\
g & \propto & f_{g}(\vartheta), \label{cumulant_fss}
\end{eqnarray}
where $\beta=1/8$, $\gamma=7/4$, $\alpha=0$ (logarithmic divergence)
and $\nu=1$ are the critical exponents (the Ising 2d ones). The
$\hat{\alpha}$, $\hat{\beta}$, $\hat{\gamma}$ and $\hat{\lambda}$
are the logarithmic correction exponents. The $f_{i}(\vartheta)$ are
the FSS functions with a logarithmic corrected scaling variable
\begin{equation}
\vartheta=L^{1/\nu}\left(T-T_{c}\right)\left| \ln \left| T-T_{c}\right|\right|^{-\hat{\lambda}}.
\end{equation}
The correction exponents $\hat{\alpha}$, $\hat{\beta}$, $\hat{\gamma}$ and $\hat{\lambda}$ obey the following scaling relations\cite{kenna:2012}
\begin{equation}
\hat{\alpha} = 1-d\nu\hat{\lambda}
\label{fss_scaling_relations1}
\end{equation}
\begin{equation}
2 \hat{\beta} - \hat{\gamma} = -d\nu\hat{\lambda},
\label{fss_scaling_relations2}
\end{equation}
where $d$ is the dimensionality of the system. The scaling relation
(\ref{fss_scaling_relations1}) is valid only for $\alpha = 0$
(logarithmic divergences), in the general case, $\hat{\alpha} =
-d\nu\hat{\lambda}$. For  $\alpha = 0$ and $\hat{\alpha} = 0$, we
have the double logarithmic divergence ($\ln \ln L$) of the specific
heat as seen for the 2d diluted Ising model\cite{kenna:2012}.

We used $1 \times 10^{5}$ MCM (Monte-Carlo Markov) steps to make the
$N_t=600$ system replicas (each system replica has a different
temperature) reach the equilibrium state and the independent
steady-state system configurations are estimated in the next $2
\times 10^{6}$ MCM steps with $10$ MCM steps between one system
state and another one to avoid self-correlation effects. Every MCM
steps are composed of two parts, a sweep, and a swap. One sweep is
accomplished when all N spins were investigated if they flip or not
and one swap is accomplished if all the $N_t$ lattices are
investigated if they exchange or not their temperatures (swap part).
We carried out $10^{5}$ independent steady-state configurations to
calculate the needed thermodynamic averages.

\section{Results and Discussion}

\begin{figure}[h!]
\begin{center}
\includegraphics[scale=0.5]{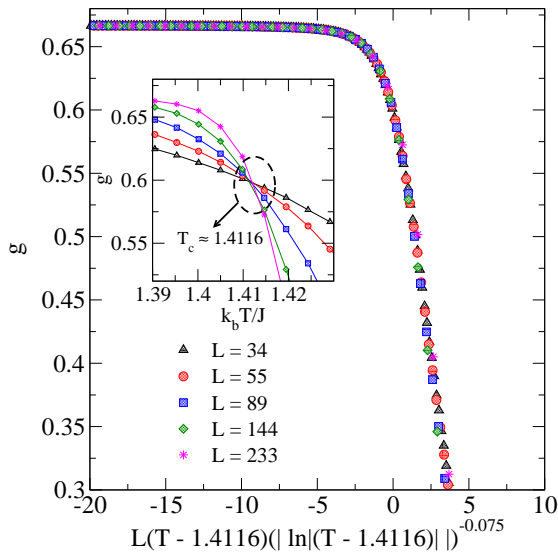}
\end{center}
\caption{(Color Online) Data collapse of the Binder Cumulant $g$
versus the scaling parameter $L^{1/\nu}(T-T_c)\left|
\ln\left|T-T_c\right| \right|^{-\hat{\lambda}}$ for different
lattice sizes $L$ where we considered the modulation C)
of the triangular lattice (discussed in the text).
Inset: Binder Cumulant versus the temperature for different lattice
sizes. The values of $L$ obey the Fibonacci sequence. We estimated
the critical temperature $T_{c}\approx 1.4116$, as shown in the
inset, by averaging the numerical values of the temperatures where
the curves intersect each other, identified by a circle with a
dashed line. The best collapse was done by using the logarithmic
correction exponent $\hat{\lambda}=0.075$. The model is in the Ising
universality class with logarithmic corrections.}
\label{fig_binder_cumulant}
\end{figure}

We focus on the modulation scheme C) in all results with the exception of
specific heat because of its distinctive critical behavior as we will show
in the following. First, we estimate the critical temperature by using the
Binder cumulant $g$ given by Eq. (\ref{binder_cumulant}). We show
the Binder cumulant in the inset of Fig.(\ref{fig_binder_cumulant})
for the modulation scheme C). The critical temperature $T_{c}$
is estimated at the point where the curves for different size
lattices intercept each other. From Fig.(\ref{fig_binder_cumulant}),
we obtained $T_{c} \approx 1.4116$ for the modulation C). For modulations
A) and B), we obtained $T_{c} \approx 2.6284$ and $T_{c} \approx 2.1540$,
respectively.

\begin{figure}[h!]
\begin{center}
\includegraphics[scale=0.5]{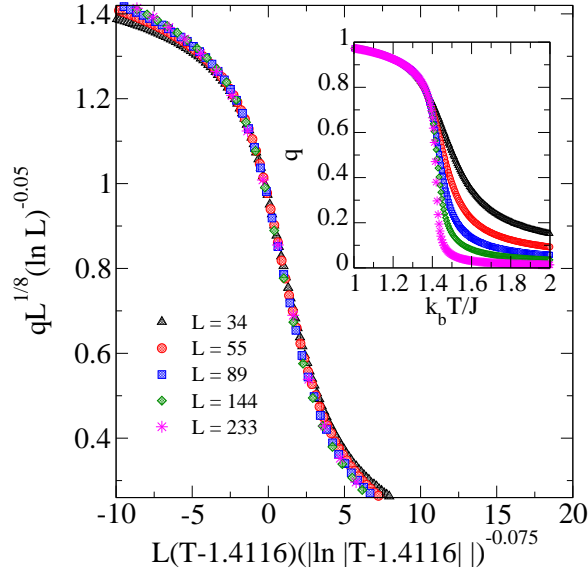}
\end{center}
\caption{(Color Online) Data collapse of the order parameter $q$,
rescaled by $L^{\beta/\nu}\left(\ln
L\right)^{\hat{\beta}+\beta\hat{\lambda}}$ versus the scaling
parameter $L^{1/\nu}(T-T_c)\left| \ln\left|T-T_c\right|
\right|^{-\hat{\lambda}}$ for different lattice sizes $L$. Here, we considered
the modulation scheme C) (discussed in text). Inset:
order parameter $q$ as a function of temperature $T$ for different
lattice sizes $L$. The values of $L$ obey the Fibonacci sequence.
The curves suggest a second order phase transition. The best
collapse is done by using the values for the logarithmic correction
exponents: $\hat{\beta} = -0.06$ and $\hat{\lambda}=0.075$. The
model is in the Ising universality class with logarithmic
corrections.}
\label{fig_orderparameter}
\end{figure}

Next we show the order parameter behavior $q$ as a function of
temperature $T$, for the modulation scheme C). The result of the inset in the
Fig.(\ref{fig_orderparameter}) suggests the presence of a
second-order phase transition in the system. Also, the
Fig.(\ref{fig_orderparameter}) presents the data collapse using the
FSS with logarithmic relation written on Eq.(\ref{q_fss}). The best
collapse was obtained by ising the values for the critical exponents
$\nu=1$, $\beta=1/8$ and the correction exponents
$\hat{\beta}=-0.06$ and $\hat{\lambda}=0.075$.

\begin{figure}[h!]
\begin{center}
\includegraphics[scale=0.5]{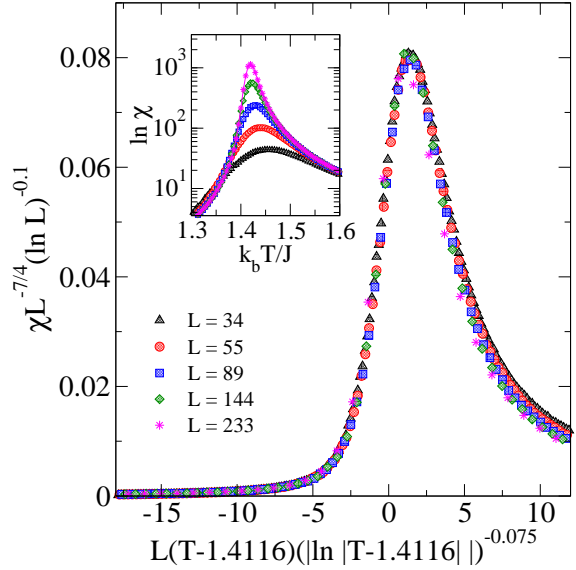}
\end{center}
\caption{(Color Online) Data collapse of the susceptibility $\chi$,
rescaled by $L^{-\gamma/\nu}\left(\ln
L\right)^{\hat{\gamma}-\gamma\hat{\lambda}}$ versus the scaling
parameter $L^{1/\nu}(T-T_c)\left| \ln\left|T-T_c\right|
\right|^{-\hat{\lambda}}$ for different lattice sizes $L$. Here, we considered
the modulation scheme C) (discussed in text). Inset:
Susceptibility $\chi$ as a function of temperature $T$ for different
lattice sizes $L$. The values of $L$ obey the Fibonacci sequence.
The susceptibility diverges at $T_{c}$ in the large lattice size
limit suggesting a second order phase transition. The best collapse
is done by using the values for the logarithmic correction
exponents: $\hat{\gamma} = 0.03$ and $\hat{\lambda}=0.075$. The
model is in the Ising universality class with logarithmic
corrections.} \label{fig_susceptibility}
\end{figure}

Continuing the analysis of the behavior critical of the system, the
inset of the Fig.(\ref{fig_susceptibility}) show the susceptibility
$\chi$ as a function of temperature $T$ for the modulation scheme C).
In the large lattice size limit, the susceptibility diverges
at $T_{c}\approx 1.4116$. The Fig.(\ref{fig_susceptibility}) also show
the data collapse of the susceptibilities for different lattice sizes
according to FSS with logarithmic correction relation given in the
Eq.(\ref{susceptibility_fss}). All maxima are well fitted by using
the scale relation with logarithmic correction ($\hat{\gamma}=0.03$
and $\hat{\lambda}=0.075$) and the 2D Ising critical exponents $\gamma=1.75$ and $\nu=1$.

\begin{figure}[h!]
\begin{center}
\includegraphics[scale=0.5]{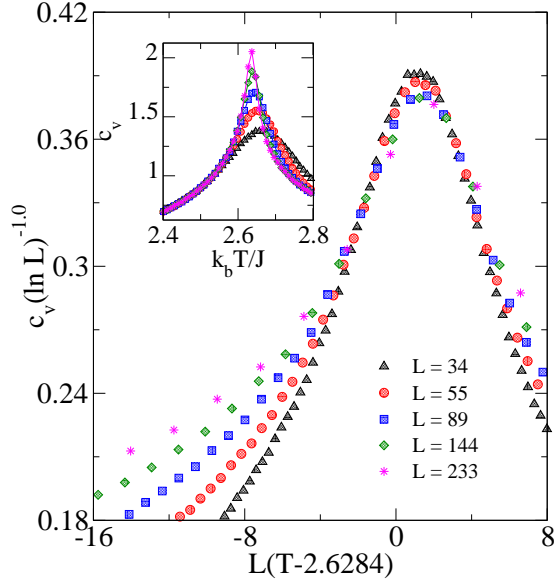}
\end{center}
\caption{(Color Online) Specific Heat $c$, rescaled by $1 / \ln L$
versus the scaling parameter $L^{1/\nu}(T-T_c)$ for different lattice
sizes $L$ and for modulation scheme A) of the triangular lattice.
Inset: Specific Heat $c$ as a function of temperature $T$ for different
lattice sizes $L$. The values of $L$ obey the Fibonacci sequence. We
can see that the usual FSS relation without logarithmic corrections does
collapse our numerical data obtained from the simulations of the
triangular lattice with modulation scheme A).}
\label{fig_specificheat-nolncorrection-one-interactions}
\end{figure}

\begin{figure}[h!]
\begin{center}
\includegraphics[scale=0.5]{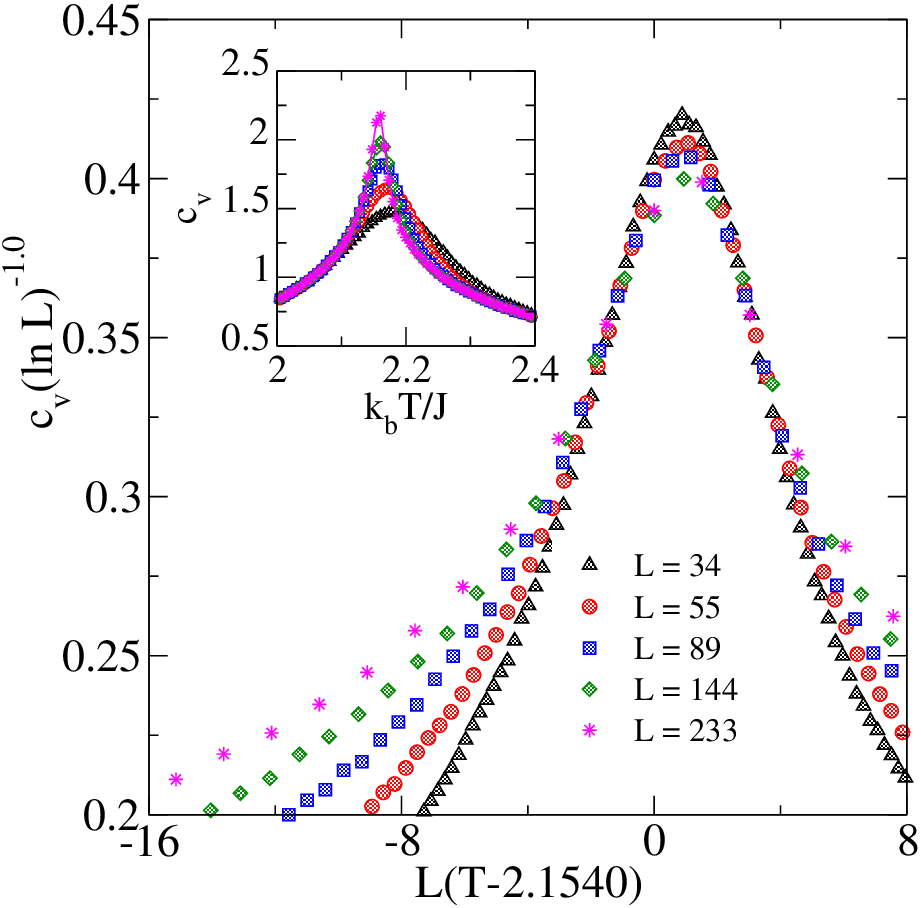}
\end{center}
\caption{(Color Online) Specific Heat $c$, rescaled by $1 / \ln L$
versus the scaling parameter $L^{1/\nu}(T-T_c)$ for different lattice
sizes $L$ and for modulation scheme B) of the triangular lattice.
Inset: Specific Heat $c$ as a function of temperature $T$ for different
lattice sizes $L$. The values of $L$ obey the Fibonacci sequence. We can
see that the usual FSS relation without logarithmic corrections does collapse
our numerical data obtained from the simulations of the triangular
lattice with modulation scheme B).}
\label{fig_specificheat-nolncorrection-two-interactions}
\end{figure}

Finally, we show the specific heat $c$, given by the Eq.(\ref{specific_heat}), at the inset of the
Figs.(\ref{fig_specificheat-nolncorrection-one-interactions}),
(\ref{fig_specificheat-nolncorrection-two-interactions}) for the modulation schemes A) and B),
respectively. We see that the usual scaling relation with a
pure logarithm divergence fits all data, being consistent with no
logarithm corrections for modulation schemes A) and B). We note in the Figs.
(\ref{fig_specificheat-nolncorrection-one-interactions}) and
(\ref{fig_specificheat-nolncorrection-two-interactions}) for
modulation schemes A) and B), the maxima of the specific heat scales
with $1/\ln L$, like the Ising ferromagnetic model.

\begin{figure}[h!]
\begin{center}
\includegraphics[scale=0.5]{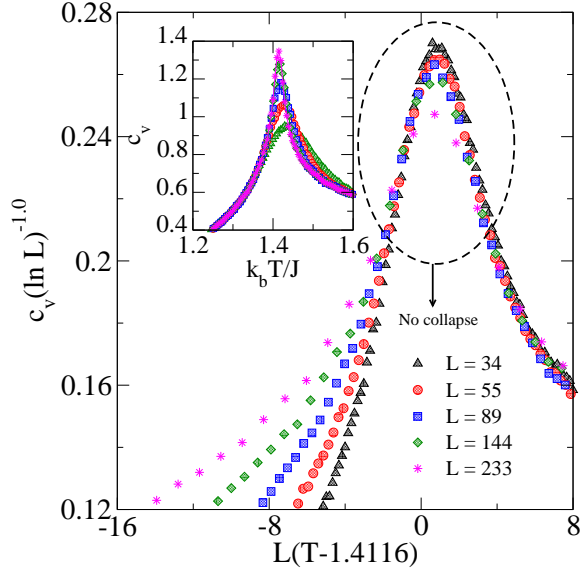}
\end{center}
\caption{(Color Online) Specific Heat $c$, rescaled by $1 / \ln L$ versus the
scaling parameter $L^{1/\nu}(T-T_c)$ for different lattice sizes $L$ and for
modulation scheme C) of the triangular lattice. Inset: Specific Heat $c$ as a
function of temperature $T$ for different lattice sizes $L$. The values of $L$
obey the Fibonacci sequence. We can see that the FSS relation without logarithmic
corrections does not collapse our numerical data with three modulations, as
is identified by a circle with dashed line.}
\label{fig_specificheat-nolncorrection}
\end{figure}

\begin{figure}[h!]
\begin{center}
\includegraphics[scale=0.5]{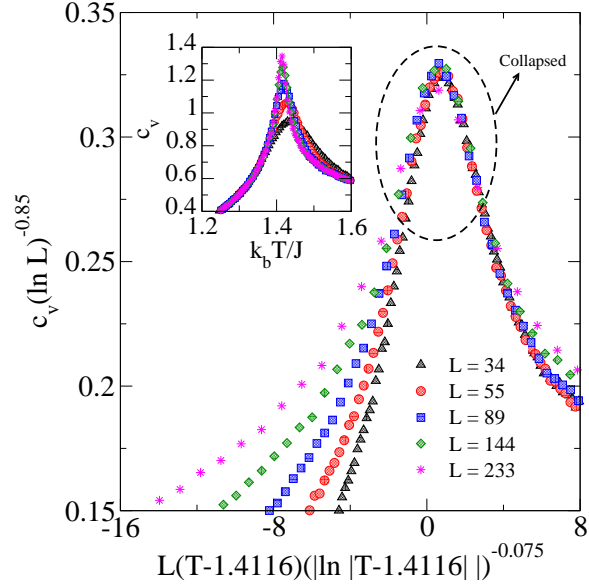}
\end{center}
\caption{(Color Online) Data collapse of the Specific Heat $c$, rescaled
by $\left( \ln L \right) ^{-\hat{\alpha}}$ versus the scaling parameter
$L^{1/\nu}(T-T_c)\left| \ln\left|T-T_c\right| \right|^{-\hat{\lambda}}$
for different lattice sizes $L$ and for modulation scheme C) of the triangular
lattice. Inset: Specific Heat $c$ as a function of temperature $T$ for
different lattice sizes $L$. The values of $L$ obey the Fibonacci sequence.
The best collapse is done by using the values for the logarithmic correction
exponents: $\hat{\alpha} = 0.85$ and $\hat{\lambda}=0.075$, as is identified
by a circle with dashed line. The model with modulation scheme C) is in the
Ising universality class with logarithmic corrections.}
\label{fig_specificheat}
\end{figure}

However, the modulation scheme C) have a different critical behavior from the
other previous cases, as we anticipated. The maxima of the specific heat $c$ do not
collapse with the $1/\ln L$, as emphasized by a circle with the
dashed line in the Fig. (\ref{fig_specificheat-nolncorrection}).
Therefore, using the scaling relation without logarithmic
corrections does not collapse our numerical data for the specific
heat $c$. However, we obtain a good collapse by using the scaling
relations written on Eq.(\ref{fss_scaling_relations1}) and our best
estimates for the logarithm correction exponents are $\hat{\alpha} =
0.85$ and $\hat{\lambda} = 0.075$ which obeys the scaling relations
for the logarithmic correction exponents given in
Eq.(\ref{fss_scaling_relations1}). The Fig.(\ref{fig_specificheat})
show the collapse of the maxima of the specific heat, as indicated
by the circle with the dashed line.

\section{Conclusions}

We have presented a theoretical model on a triangular lattice with
quasiperiodic long-ranged order based on Fibonacci quasiperiodic
sequence, with competing interactions, and we have obtained a
critical behavior of a second order phase transition, driven by the
temperature at the triangular lattice modulated by three different
ways A), B), and C), dependent on the number of interactions in each lattice
plaquette is changed according to the quasiperiodic sequence.

Note that the three modulating schemes A), B) and C) can be sorted
as in increasing order of modulating strenght, where we modulate an
increasing number of lattice bonds and introduce increasing corrections on
partition function in order to observe a change in the critical behavior.
In fact, modulation scheme C) is sufficient to introduce logarithm
corrections in the thermodynamic properties. This is a signal of
a marginal behavior depending on the modulation, placed between the case
where we have a Ising ferromagnetic long-range order at lower temperatures
and the situation for a sufficiently strong modulation where we have
only a paramagnetic phase for any finite temperature. In the marginal
modulation we have a ferrimagnetic ordering and logarithm corrections
in the critical behavior.

In the case when we choose only one interaction and two interactions
to be modulated, the system obeys the same critical behavior of the
pure Ising ferromagnetic model, as shown in Figs.
(\ref{fig_specificheat-nolncorrection-one-interactions}) and
(\ref{fig_specificheat-nolncorrection-two-interactions}) where the specific heat
have a logarithm divergence. However, the case with the three
plaquette interactions chosen to be modulated, the system deviated
from the pure model where the system retains the same
universality class, but with logarithm corrections in its critical
behavior. In this case, at the low-temperature limit, we obtained an
aperiodic ferrite phase with critical temperature
$T_{c}\approx 1.4116$, which is different from the critical temperatures of
the model on the square lattice, modulated by using
Fibonacci\cite{PhysRevE.89.042139} and Octonacci(\cite{jstat.2017.123302})
sequences.

Specifically, we have obtained the critical exponents $\beta=1/8$,
$\gamma=7/4$ and $\nu=1$ (Ising universality class) and the
estimates for logarithmic correction exponents given by
$\hat{\alpha}=0.85$, $\hat{\beta}=-0.06$, $\hat{\gamma}=0.03$ and
$\hat{\lambda}=0.075$ in the case of equal antiferromagnetic and
ferromagnetic strengths. The critical exponents of the logarithmic
correction to the triangular lattice is not the same as the square
lattice. Therefore, the quasiperiodic ordering is marginal in the
sense of introducing logarithmic corrections as in seen for 4-state
2D Potts model, Fibonacci sequence\cite{PhysRevE.89.042139}
and Octonacci sequence (\cite{jstat.2017.123302})
in the square lattice.

\section*{Acknowledgments}

We would like to thank CNPq (Conselho Nacional de Desenvolvimento
Cient\'{\i}fico e tecnol\'{o}gico) and FAPEPI (Funda\c{c}\~{a}o de
Amparo a Pesquisa do Estado do Piau\'{\i}) for the financial support.
We acknowledge the use of Dietrich Stauffer Computational Physics
Lab - UFPI, Teresina, Brazil where the numerical simulations were
performed.

\bibliography{textv_PRE}

\end{document}